\documentclass{article}

\PassOptionsToPackage{numbers,sort&compress}{natbib}
\usepackage[main,final]{neurips_2026}

\usepackage[utf8]{inputenc}
\usepackage[T1]{fontenc}
\usepackage{mathtools,amssymb,bm,empheq,nicefrac,upgreek,siunitx}
\usepackage{url}
\usepackage{booktabs}
\usepackage{amsfonts}
\usepackage[nopatch=footnote]{microtype}
\usepackage{cite}
\usepackage[caption=false,labelformat=parens,subrefformat=parens]{subfig}

\usepackage{graphicx}
\usepackage{tabularx}
\usepackage{array}
\usepackage{enumitem}
\usepackage{listings}
\usepackage{xcolor}
\usepackage[section]{placeins}

\definecolor{linkblue}{RGB}{20,90,255}

\definecolor{uolblue}{RGB}{0,62,107}       

\definecolor{blue1}{RGB}{0,78,159}        
\definecolor{blue2}{RGB}{0,171,217}       
\definecolor{blue3}{RGB}{91,197,242}      
\definecolor{blue4}{RGB}{161,217,248}     

\definecolor{green1}{RGB}{0,120,120}      
\definecolor{green2}{RGB}{0,168,121}      
\definecolor{green3}{RGB}{148,193,28}     
\definecolor{green4}{RGB}{199,211,0}      

\definecolor{orange1}{RGB}{213,59,10}     
\definecolor{orange2}{RGB}{238,113,0}     
\definecolor{orange3}{RGB}{243,145,0}     
\definecolor{orange4}{RGB}{253,195,0}     

\definecolor{lila1}{RGB}{100,89,144}      
\definecolor{lila2}{RGB}{177,152,197}     
\definecolor{lila3}{RGB}{191,117,170}     
\definecolor{lila4}{RGB}{218,145,186}     

\definecolor{VeryDarkGray}{RGB}{45,45,45}

\usepackage[
    colorlinks=true,
    linkcolor=linkblue,
    citecolor=linkblue,
    urlcolor=linkblue,
    linktoc=all
]{hyperref}
\makeatletter
\renewcommand{\NAT@open}{\textcolor{linkblue}{[}}
\renewcommand{\NAT@close}{\textcolor{linkblue}{]}}
\makeatother

\usepackage[nameinlink, capitalise]{cleveref}

\graphicspath{{Figures/}{Figures/Logos/}}

\newcolumntype{Y}{>{\raggedright\arraybackslash}X}

\setlist[itemize]{leftmargin=1.5em,itemsep=1pt,topsep=3pt}
\setlist[enumerate]{leftmargin=1.8em,itemsep=1pt,topsep=3pt}

\title{The Hearpiece Database of Binaural Room Impulse Responses and Driver Responses on KEMAR}

\author{
Tong Xiao$^{1}$,
Reinhild Roden$^{2}$,
Matthias Blau$^{2}$,
and Simon Doclo$^{1}$ \\[0.5em]
$^{1}$ Department of Medical Physics and Acoustics and Cluster of Excellence Hearing4all,\\
Carl von Ossietzky Universit\"at Oldenburg, Germany\\
$^{2}$ Institut f\"ur H\"ortechnik und Audiologie,
Jade Hochschule, Oldenburg, Germany \\[0.5em]
\texttt{tong.xiao@uni-oldenburg.de, reinhild.roden@jade-hs.de,}\\
\texttt{matthias.blau@jade-hs.de, simon.doclo@uni-oldenburg.de}
}

\begin{document}

\maketitle

\noindent
\small
\textbf{Version:} 1.0
\hfill
\textbf{Last update:} October 27, 2026

\vspace{0.0em}

\noindent
\small
This database is published under the
\href{https://creativecommons.org/licenses/by-sa/4.0/}
{Creative Commons Attribution-ShareAlike 4.0 International (CC BY-SA 4.0) license}.
You are free to share and adapt the material, provided appropriate credit is given and derivative works are distributed under the same license.

\vspace{0.3em}

\noindent
\small
\textcopyright{} Carl von Ossietzky Universit\"at Oldenburg.

\vspace{0.4em}

\noindent
\small
\textbf{Acknowledgments.} This research was funded by the Deutsche Forschungsgemeinschaft (DFG, German Research Foundation) -- Project-ID 352015383 -- SFB 1330 C1, and under Germany's Excellence Strategy -- EXC 2177/2 -- Project ID 390895286.

\vspace{0.6em}

\begin{abstract}
This document describes the measurement setup, signal processing, data organization, and MATLAB file format of a database of acoustic responses measured on a Knowles Electronics Manikin for Acoustic Research (KEMAR) using the Hearpiece, a multi-microphone, multi-driver in-ear research device. The data include binaural room impulse responses (BRIRs) and Hearpiece driver responses, together with center-microphone room impulse responses and a loopback impulse response of the playback and recording system. The BRIRs cover two Hearpiece configurations, three room conditions, three independent insertions per configuration, and 24 horizontal source directions. The database supports reproducible research on binaural and hearing-device signal processing, with potential applications in speech enhancement, spatial filtering and source separation, dereverberation, active noise and occlusion control, and acoustic feedback cancellation. The database, MATLAB loading script, and technical documentation are publicly available at \url{https://doi.org/10.5281/zenodo.23667308}.
\end{abstract}

\vspace{-1em}

\begingroup
\setlength{\parskip}{0pt}
\tableofcontents
\endgroup
\clearpage


\section{Overview}

\subsection{Introduction and scope}
\label{sec:database-scope}

The Hearpiece is a one-size-fits-all research earpiece equipped with multiple microphones and drivers for hearing-device research \citep{denk2019earpiece}. Its design, transducer configuration, and electroacoustic characteristics are described in detail by \citet{denk2019earpiece}, while additional information on the hardware, connections, microphones, and drivers is provided in the Hearpiece technical documentation \citep{hearpiece2019technical}. A complementary anechoic database of individual Hearpiece acoustic responses was published by \citet{denk2021hearpiece}. It includes head-related impulse responses (HRIRs) and driver responses.

\begin{figure}[htbp]
    \centering
    \includegraphics[width=0.5\linewidth]{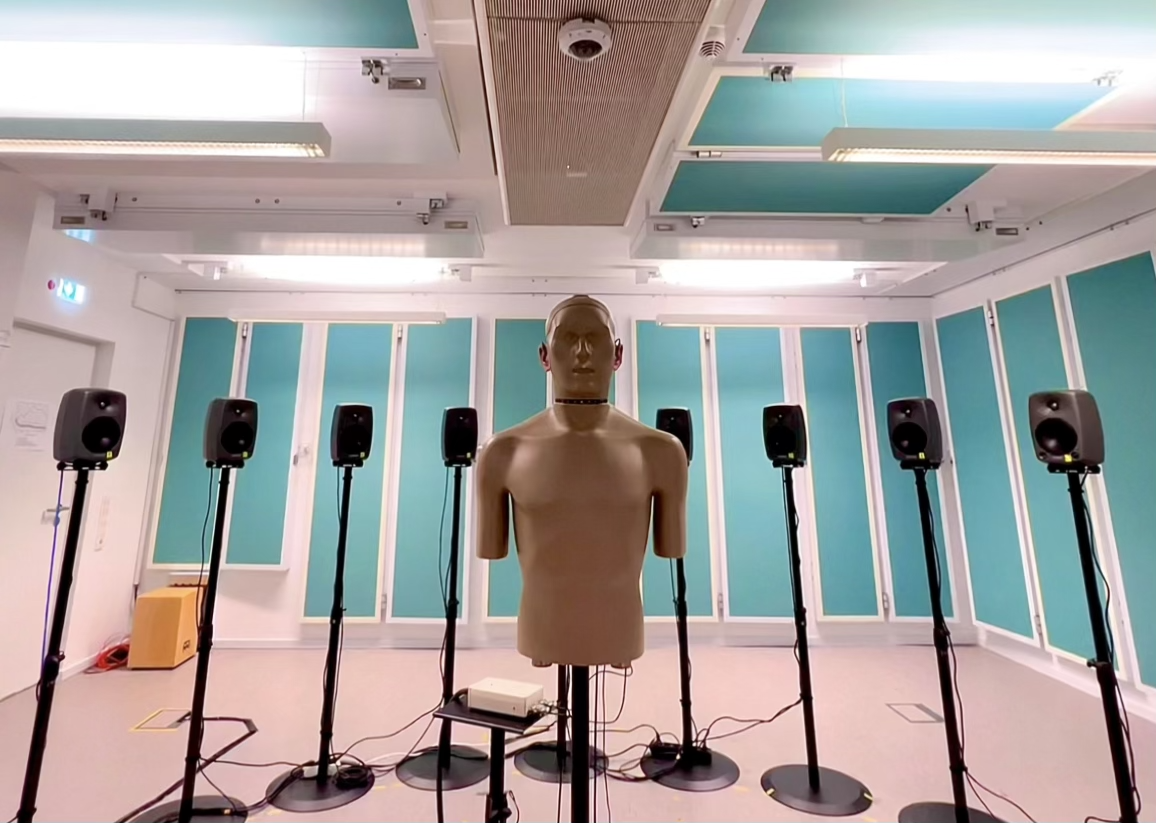}
    \vspace{0pt}
    \caption{Measurement setup with a KEMAR positioned at the center of the circular loudspeaker array.}
    \label{fig:measurement-setup}
    \vspace{1em}
\end{figure}

Building directly on the previous Hearpiece database of \citet{denk2021hearpiece}, the present database extends the established KEMAR-based measurement framework from anechoic HRIRs to room-dependent binaural room impulse responses (BRIRs), while also providing Hearpiece driver responses. Using a common measurement setup, it provides BRIRs and driver responses for two Hearpiece fitting configurations, closed and vented, represented by the devices \texttt{AC0031} and \texttt{AV0006}, respectively, with three independent insertions for each device. The BRIRs cover three room conditions and 24 horizontal source directions across 10 recording channels. This combination brings together multichannel responses at the Hearpiece microphones and Eardrum locations, controlled room variation, configuration and insertion variability, and measured driver responses within a single database. An overview of the measurement setup is shown in \cref{fig:measurement-setup}.

\FloatBarrier

The BRIR data comprise
\begin{equation}
2\;\text{configurations}
\times 3\;\text{room conditions}
\times 3\;\text{insertions}
= 18\;\text{BRIR files}.
\end{equation}
Each BRIR file contains $24 \times 10 = 240$ impulse responses, corresponding to 24 source directions and 10 recording channels. The BRIR part of the database therefore contains
\begin{equation}
18 \;\text{BRIR files} \times 24 \;\text{source directions} \times 10 \;\text{recording channels} = 4320\;\text{impulse responses}.
\end{equation}

The driver-response data contain
\begin{equation}
2\;\text{configurations}
\times 3\;\text{insertions}
= 6\;\text{driver-response files}.
\end{equation}
Each driver-response file contains $4 \times 10 = 40$ impulse responses, corresponding to four Hearpiece drivers and 10 recording channels. The driver-response portion of the database therefore contains
\begin{equation}
6 \;\text{driver-response files} \times 4 \;\text{drivers} \times 10 \;\text{recording channels} = 240\;\text{impulse responses}.
\end{equation}

In addition, the database contains center-microphone room impulse responses (RIRs) and a loopback impulse response of the playback and recording system. The center-microphone responses were measured at the center of the loudspeaker array without KEMAR present and provide a reference characterization of the three room conditions. The loopback impulse response used to compensate the BRIR and driver-response data is also included. All impulse responses are provided at a sampling rate of \qty{48}{\kilo\hertz}.

\subsection{Database organization}
\label{sec:database-organization}

The database is organized into four folders containing BRIRs, driver responses, center-microphone RIRs, and the loopback measurement, all in MATLAB MAT format. A MATLAB script, \texttt{Load\_Hearpiece\_Database.m}, is also provided to load the data for a selected Hearpiece device, room condition, and insertion into the MATLAB workspace. The complete structure is:
\begin{lstlisting}
Hearpiece_KEMAR_Room_Database/
|
|-- BRIRs_MAT/
|   |-- BRIR_KEMAR_AC0031_reverb1_1.mat
|   |-- BRIR_KEMAR_AC0031_reverb1_2.mat
|   |-- BRIR_KEMAR_AC0031_reverb1_3.mat
|   |-- BRIR_KEMAR_AC0031_reverb2_1.mat
|   |-- BRIR_KEMAR_AC0031_reverb2_2.mat
|   |-- BRIR_KEMAR_AC0031_reverb2_3.mat
|   |-- BRIR_KEMAR_AC0031_reverb3_1.mat
|   |-- BRIR_KEMAR_AC0031_reverb3_2.mat
|   |-- BRIR_KEMAR_AC0031_reverb3_3.mat
|   |-- BRIR_KEMAR_AV0006_reverb1_1.mat
|   |-- BRIR_KEMAR_AV0006_reverb1_2.mat
|   |-- BRIR_KEMAR_AV0006_reverb1_3.mat
|   |-- BRIR_KEMAR_AV0006_reverb2_1.mat
|   |-- BRIR_KEMAR_AV0006_reverb2_2.mat
|   |-- BRIR_KEMAR_AV0006_reverb2_3.mat
|   |-- BRIR_KEMAR_AV0006_reverb3_1.mat
|   |-- BRIR_KEMAR_AV0006_reverb3_2.mat
|   `-- BRIR_KEMAR_AV0006_reverb3_3.mat
|
|-- DriverResponses_MAT/
|   |-- DriverResponse_KEMAR_AC0031_1.mat
|   |-- DriverResponse_KEMAR_AC0031_2.mat
|   |-- DriverResponse_KEMAR_AC0031_3.mat
|   |-- DriverResponse_KEMAR_AV0006_1.mat
|   |-- DriverResponse_KEMAR_AV0006_2.mat
|   `-- DriverResponse_KEMAR_AV0006_3.mat
|
|-- CenterMic_MAT/
|   |-- CenterMic_reverb1.mat
|   |-- CenterMic_reverb2.mat
|   `-- CenterMic_reverb3.mat
|
|-- Calibration_MAT/
|   `-- Loopback.mat
|
|-- Load_Hearpiece_Database.m
|
`-- Documentation.pdf
\end{lstlisting}


\subsection{Related databases}

Several other publicly available acoustic databases are relevant to hearing-device research:
\begin{itemize}
    \item The \textit{Transfer Function Measurements for Simulating Environmental Noise at Hearable Microphones} contains impulse-response measurements for 18 users wearing bilateral Hearpieces, using a closed fitting and eight horizontal source directions. The measurements are limited to the In-Ear and Concha microphones and do not include Hearpiece driver responses \citep{ohlenbusch2024transfer}.

    \item The \textit{Paths for Active Noise Cancellation Development and Research (PANDAR)} database contains primary, secondary, and acoustic feedback paths measured with in-ear active noise control headphones. It includes measurements from human subjects and direction-dependent primary paths for 72 horizontal source directions on a dummy head, using one outer microphone, one inner microphone, and the internal loudspeaker \citep{liebich2019database}.

    \item The \textit{Database of Multichannel In-Ear and Behind-the-Ear Head-Related and Binaural Room Impulse Responses} provides multichannel HRIRs and BRIRs measured on a head-and-torso simulator using bilateral behind-the-ear hearing aid dummies and in-ear microphones. Measurements are available for an anechoic chamber and several natural environments, with the number of source directions and available recording channels varying between environments \citep{kayser2009database}.

    \item The \textit{Binaural Room Impulse Responses with Uniformly Distributed External Microphones (BRUDEX)} database contains acoustic impulse responses for a binaural hearing-aid setup together with 36 spatially distributed external microphones, 12 source positions, and three reverberation conditions. Its main emphasis is on spatially distributed microphone measurements for applications such as noise reduction, source localization, and dereverberation \citep{fejgin2023brudex}.

    \item The \textit{IHTA-indHARTF -- Database of Individual Behind-the-Ear Hearing-Aid-Related Transfer Functions with High Spatial Resolution} provides individual hearing-aid-related transfer functions (HARTFs) from dual-microphone behind-the-ear hearing aids and corresponding in-ear head-related transfer functions (HRTFs) for 30 participants. The measurements were obtained in a hemi-anechoic chamber and focus on individual spatial transfer functions rather than room-dependent or driver-response measurements \citep{pausch2022ihta}.
\end{itemize}


\section{Measurement setup}
\label{sec:measurement-setup}

\subsection{Room conditions}
\label{sec:room-conditions}

Measurements were conducted in the {Variable Acoustics Laboratory} at the Carl von Ossietzky Universit\"at Oldenburg with approximate dimensions of $7 \times 6 \times 2.7$~\si{\meter}. Three room conditions were measured and identified as \texttt{reverb1}, \texttt{reverb2}, and \texttt{reverb3}, ordered from least to most reverberant.

The room condition was varied by changing the amount of acoustic absorption, primarily by reconfiguring sound-absorbing panels on the walls, as illustrated in \cref{fig:measurement-setup}. BRIRs were measured in all three room conditions, whereas the driver responses were measured only in \texttt{reverb1}. Center-microphone RIRs are provided for all three conditions.

\subsection{KEMAR and Hearpiece}
\label{sec:kemar-hearpiece}

Measurements were performed using the same G.R.A.S. KEMAR 45BB-12 manikin used for the Hearpiece measurements reported by \citet{denk2021hearpiece}. 
KEMAR was positioned at the center of the circular loudspeaker array, facing the loudspeaker at \qty{0}{\degree} azimuth, with its interaural center approximately \qty{1.6}{\meter} above the floor. The interaural center was vertically aligned with the loudspeaker centers, defined as the midpoints between the woofer and tweeter of each loudspeaker. The KEMAR was equipped with G.R.A.S. 43BB-1 low-noise ear simulators and G.R.A.S. KB5000/5001 anthropometric pinnae, which were also used in the previous Hearpiece measurements \citep{denk2019earpiece}. For all BRIR and driver-response measurements, medium-sized silicone domes were used.

A pair of Hearpieces was fitted bilaterally to KEMAR for the BRIR and driver-response measurements. Two Hearpiece devices, representing the two fitting configurations, were measured:

\begin{itemize}
    \item \texttt{AC0031}: closed fitting,
    \item \texttt{AV0006}: vented fitting.
\end{itemize}

Each Hearpiece contains four Knowles SPH1642HT5H-1 microelectromechanical systems (MEMS) microphones \citep{knowlesSPH1642HT5H1}. The In-Ear and Outer Vent microphones are located near the inner and outer ends of the vent, respectively. The Entrance and Concha microphones are mounted on the faceplate of the Hearpiece, with the Entrance microphone located near the ear-canal entrance and the Concha microphone farther back in the cavum conchae. These microphone-location names are used throughout the database.

\begin{figure}[ht]
    \centering
    \captionsetup[subfloat]{captionskip=0pt,farskip=-10pt}
    \subfloat[]{\includegraphics[height=0.24\linewidth]{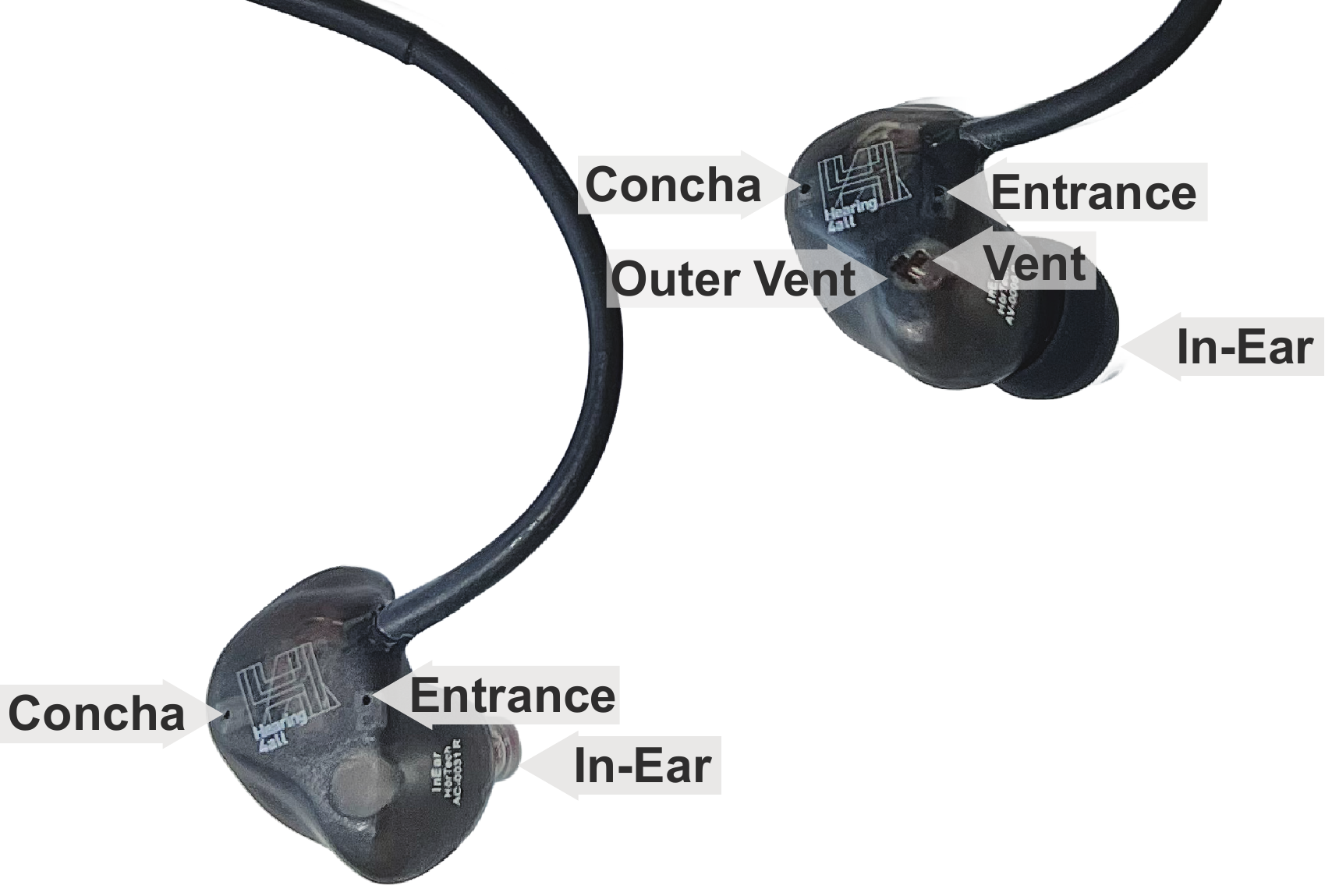}\label{fig:hearpiece-configurations}}
    \hspace{0.06\linewidth}
    \subfloat[]{\includegraphics[height=0.28\linewidth]{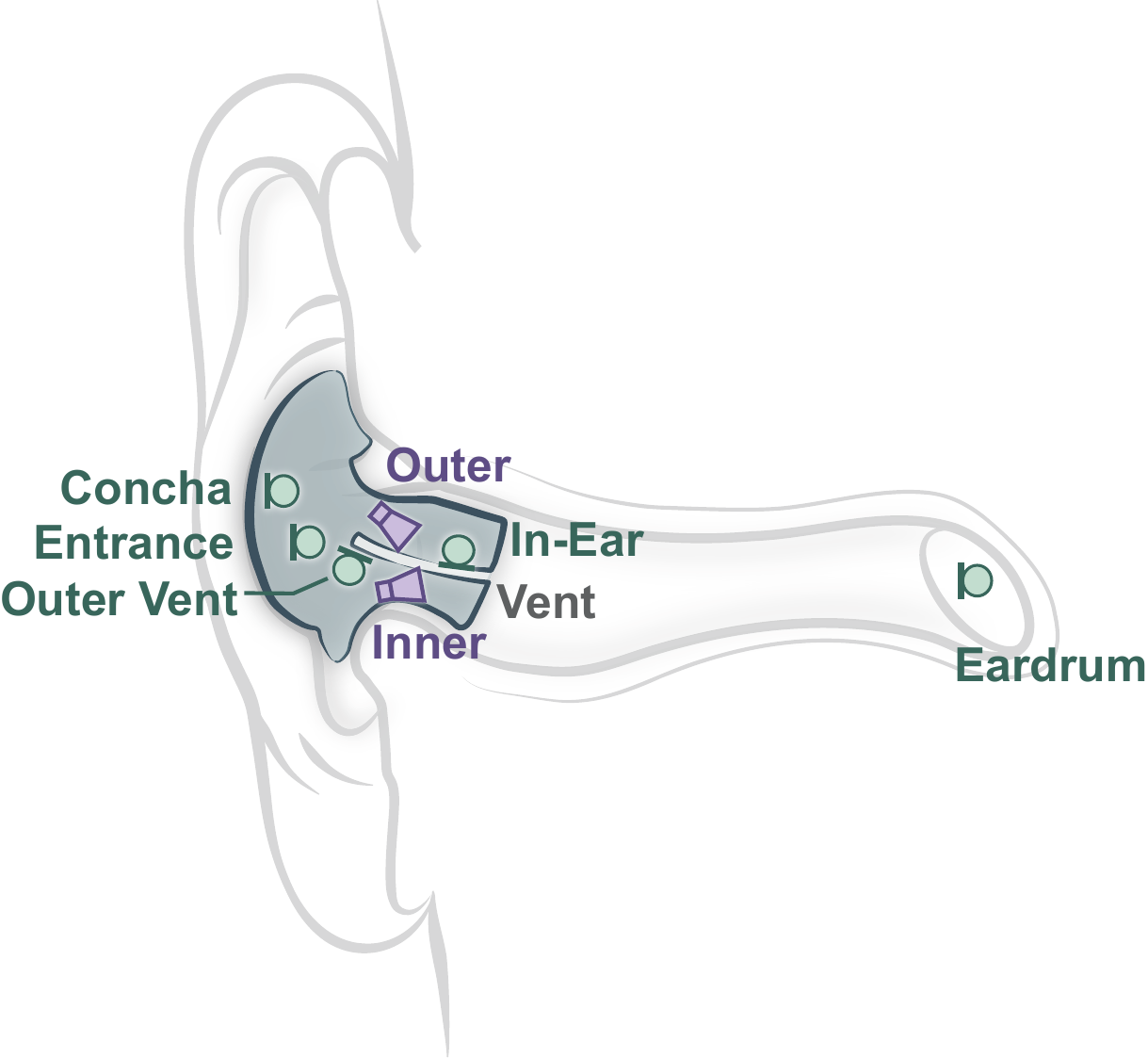}\label{fig:hearpiece-layout}}
    \vspace{-4pt}
    \caption{(a) Hearpiece configurations used in the measurements: closed fitting (\texttt{AC0031}) on the left and vented fitting (\texttt{AV0006}) on the right, with selected microphone locations indicated. (b) Schematic layout of the Hearpiece in the ear, showing the microphone locations, Inner and Outer Drivers, the vent, and the KEMAR ear-simulator recording location. }
    \label{fig:hearpiece}
    \vspace{0em}
\end{figure}

Each Hearpiece also contains two separately driven balanced-armature (BA) drivers, referred to as the Inner Driver and Outer Driver according to their positions in the device \citep{denk2019earpiece}. The Inner Driver is a Knowles WBFK-30042 used primarily for high-frequency reproduction, while the Outer Driver is a Knowles FK-26768 used primarily for lower-frequency reproduction. The Hearpiece configurations and the physical arrangement of the microphones and drivers are illustrated in \cref{fig:hearpiece}.

For each Hearpiece configuration, three independent insertions were measured. For each insertion, BRIRs were measured under all three room conditions by reconfiguring the sound-absorbing panels on the walls, without removing or repositioning the Hearpieces. The Hearpieces were removed and reinserted between successive insertions.

\FloatBarrier

\subsection{Loudspeaker array}
\label{sec:loudspeaker-array}

The playback system consisted of 24 Genelec 8030 loudspeakers arranged in a horizontal circular array of approximately \qty{2}{\meter} radius around the measurement position, with the loudspeaker centers at a height of approximately \qty{1.6}{\meter}. The loudspeakers were uniformly spaced at \qty{15}{\degree} intervals. Playback and recording were performed using an RME MADIface XT audio interface together with RME ADI-8 QS and RME OctaMic XTC converters.

\begin{figure}[ht]
    \centering
    \includegraphics[width=0.36\linewidth]{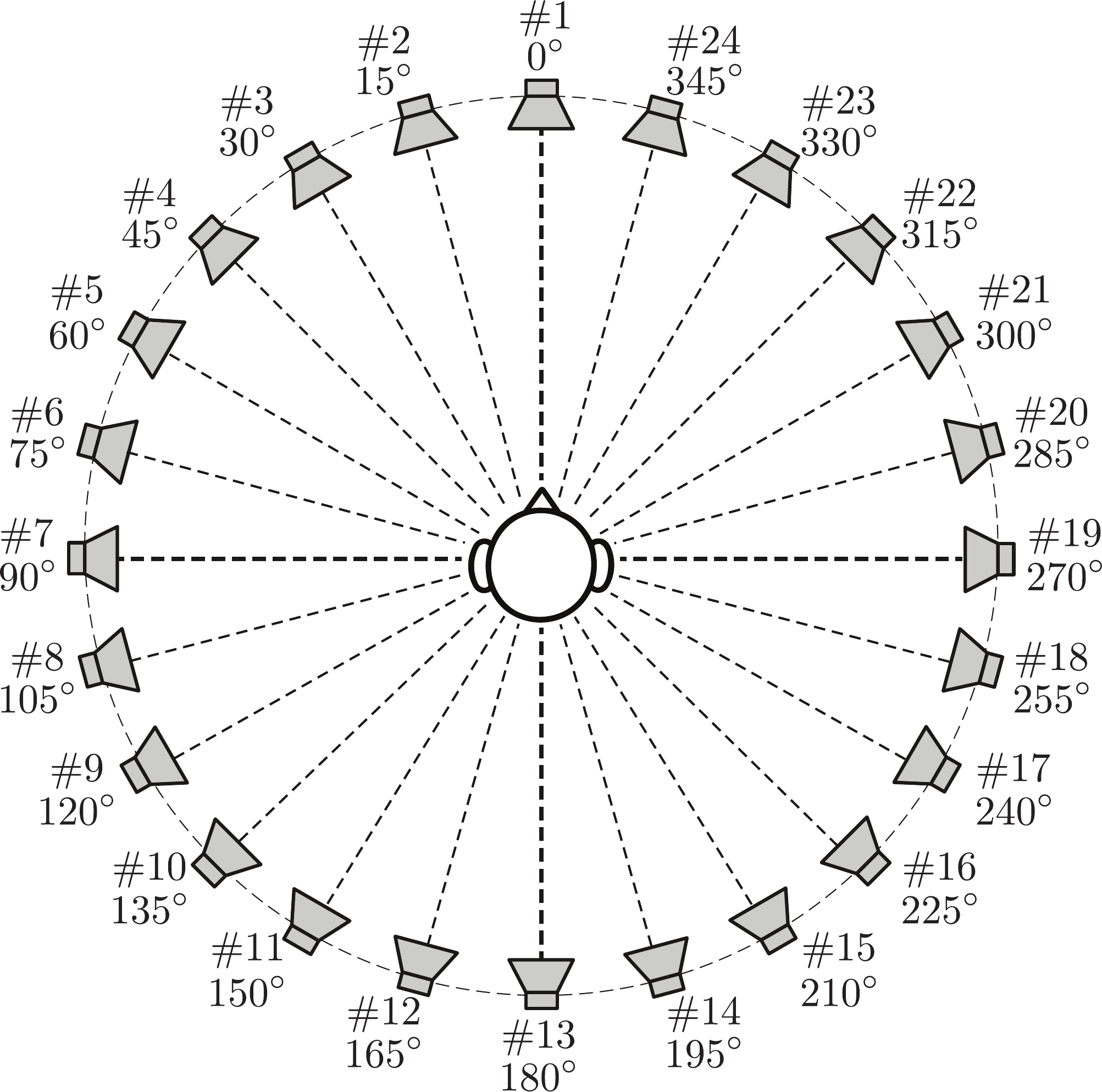}
    \vspace{0pt}
    \caption{Top view of the 24-loudspeaker array, showing the loudspeaker indexing and corresponding source azimuths used in the database.}
    \label{fig:loudspeaker-array}
    \vspace{1.0em}
\end{figure}

Each loudspeaker defines one source direction in the BRIR and center-microphone data. As illustrated in \cref{fig:loudspeaker-array}, the 24 loudspeakers are indexed from \texttt{\#1} to \texttt{\#24}, starting with \texttt{\#1} at an azimuth of \qty{0}{\degree} and increasing in \qty{15}{\degree} steps counterclockwise when viewed from above. Thus, \qty{90}{\degree} is to the left, \qty{180}{\degree} at the back, and \qty{270}{\degree} to the right. All source directions have an elevation of \qty{0}{\degree}. 

\FloatBarrier

\subsection{Recording channels}
\label{sec:recording-channels}

Both the BRIR and driver-response data contain 10 recording channels, following the terminology of the previous Hearpiece database~\citep{denk2021hearpiece}. Channels 1 and 2 are the outputs of the left and right KEMAR ear simulators and are referred to in the database as Eardrum Left and Eardrum Right. Channels 3--10 correspond to four Hearpiece microphone locations measured bilaterally: In-Ear, Entrance, Outer Vent, and Concha. The locations of the Hearpiece microphones are illustrated in \cref{fig:hearpiece}. The recording-channel ordering is given in \Cref{tab:recording-channels}.

\begin{table}[ht]
\caption{Recording-channel ordering used for the BRIR and driver-response data.}
\label{tab:recording-channels}
\centering
\begin{tabularx}{\linewidth}{@{}cY@{}}
\toprule
Channel & Recording-channel name \\
\midrule
1  & Eardrum Left \\
2  & Eardrum Right \\
3  & In-Ear Microphone Left \\
4  & In-Ear Microphone Right \\
5  & Entrance Microphone Left \\
6  & Entrance Microphone Right \\
7  & Outer Vent Microphone Left \\
8  & Outer Vent Microphone Right \\
9  & Concha Microphone Left \\
10 & Concha Microphone Right \\
\bottomrule
\end{tabularx}
\end{table}

\FloatBarrier

\subsection{Measurement signals and impulse-response estimation}
\label{sec:measurement-signals}

Impulse responses were measured using the exponential swept-sine (ESS) technique \citep{farina2000swept,novak2010nonlinear} at a sampling rate of \qty{48}{\kilo\hertz}. The sweep had a nominal duration of \qty{5}{\second} and covered the frequency range from \qty{20}{\hertz} to \qty{24}{\kilo\hertz}. The sweep was followed by \qty{2}{\second} of silence.

Each measurement was repeated three times. The recorded signals from the second and third repetitions were averaged, while the first repetition was discarded. The averaged recorded sweep response was then convolved with the corresponding regularized inverse sweep to obtain the impulse response. The causal linear impulse response was separated from the nonlinear-response components.

\subsection{Calibration and response correction}
\label{sec:calibration}

The two Eardrum channels were corrected for the relative sensitivity difference between the left and right KEMAR ear simulators using the unit-specific sensitivities from their calibration certificates \citep{grasCalibration}. These sensitivities were used only for relative left--right scaling and not for absolute pressure calibration. No frequency-response equalization was applied to these channels.

The Hearpiece microphone channels were corrected for the nominal frequency-response shape of the Knowles SPH1642HT5H-1 microphone \citep{knowlesSPH1642HT5H1}. The manufacturer's normalized response was used for this correction.

The BRIR and driver-response data were additionally compensated for the playback and recording system using the measured loopback impulse response described in \Cref{sec:loopback-measurement}. No loopback compensation or microphone-response correction was applied to the center-microphone responses. The BRIR and driver-response amplitudes should not be interpreted as absolute acoustic pressure, absolute electrical gain, or sound pressure level.


\section{Database contents and data format}
\label{sec:database-contents}

The following subsections document the four data types in the database. The BRIR files are described in \cref{sec:brir-data}, the driver-response files in \cref{sec:driver-responses}, the center-microphone RIR files in \cref{sec:centermic}, and the loopback measurement in \cref{sec:loopback-measurement}. For each data type, the file naming, organization of the impulse-response data, and metadata variables are described, with representative plots illustrating the contents of the files.

\subsection{Binaural room impulse responses}
\label{sec:brir-data}

The BRIR data are stored in \texttt{BRIRs\_MAT/}. The folder contains 18 MAT files corresponding to two Hearpiece configurations, three room conditions, and three independent insertions. The files follow the naming convention
\begin{lstlisting}
BRIR_KEMAR_<deviceID>_<reverbCondition>_<insertion>.mat
\end{lstlisting}
where \texttt{deviceID} is either \texttt{AC0031} or \texttt{AV0006}, \texttt{reverbCondition} is \texttt{reverb1}, \texttt{reverb2}, or \texttt{reverb3}, and \texttt{insertion} is 1, 2, or 3. 
\vspace{1em}

The variables contained in each BRIR MAT file are summarized in \Cref{tab:brir-vars}. Each file contains the impulse responses in \texttt{M\_data} with dimensions $\num{48000} \times 24 \times 10$, corresponding to samples $\times$ source directions $\times$ recording channels. The rows of \texttt{M\_dirs\_sph} correspond to the source-direction dimension of \texttt{M\_data}, and the entries of \texttt{c\_inchannel\_names} correspond to the recording-channel dimension. The variable \texttt{n\_delay\_common} follows the timing convention used in the previous Hearpiece database \citep{denk2021hearpiece} and specifies the number of samples preceding the nominal time origin of the stored impulse responses. In the present database, \texttt{n\_delay\_common} is set to 0 for all BRIR files, indicating that no samples are designated as negative-time samples. Calibration and response corrections applied to the BRIR data are described in \Cref{sec:calibration}.

\begin{table}[ht]
\caption{Variables contained in each BRIR MAT file.}
\label{tab:brir-vars}
\centering
\begin{tabularx}{\linewidth}{@{}l l Y@{}}
\toprule
Variable & Size/type & Description \\
\midrule
\texttt{M\_data} & $\num{48000} \times 24 \times 10$ & Impulse responses with dimensions of samples $\times$ source directions $\times$ recording channels. \\
\texttt{M\_dirs\_sph} & $24 \times 2$ & Source directions given as azimuth and elevation in degrees. \\
\texttt{c\_inchannel\_names} & $1 \times 10$ cell array & Recording-channel names. \\
\texttt{n\_delay\_common} & scalar & Number of negative-time samples preceding the nominal time origin; equal to 0. \\
\texttt{srate} & scalar & Sampling rate in Hz; equal to \num{48000}. \\
\texttt{s\_subjectID} & character array & Subject identifier; \texttt{KEMAR}. \\
\texttt{s\_deviceID} & character array & Hearpiece device identifier; \texttt{AC0031} or \texttt{AV0006}. \\
\texttt{s\_dataset} & character array & Dataset identifier. \\
\texttt{s\_reverbCondition} & character array & Room-condition identifier; \texttt{reverb1}, \texttt{reverb2}, or \texttt{reverb3}. \\
\texttt{s\_fitting} & character array & Hearpiece fitting; \texttt{closed} or \texttt{vented}. \\
\texttt{n\_insertion} & scalar & Hearpiece insertion number. \\
\texttt{c\_readme} & cell array & Brief description of the data, measurement configuration, and processing. \\
\bottomrule
\end{tabularx}
\end{table}
\FloatBarrier

\begin{figure}[ht]
    \centering
    \includegraphics[width=0.95\linewidth]{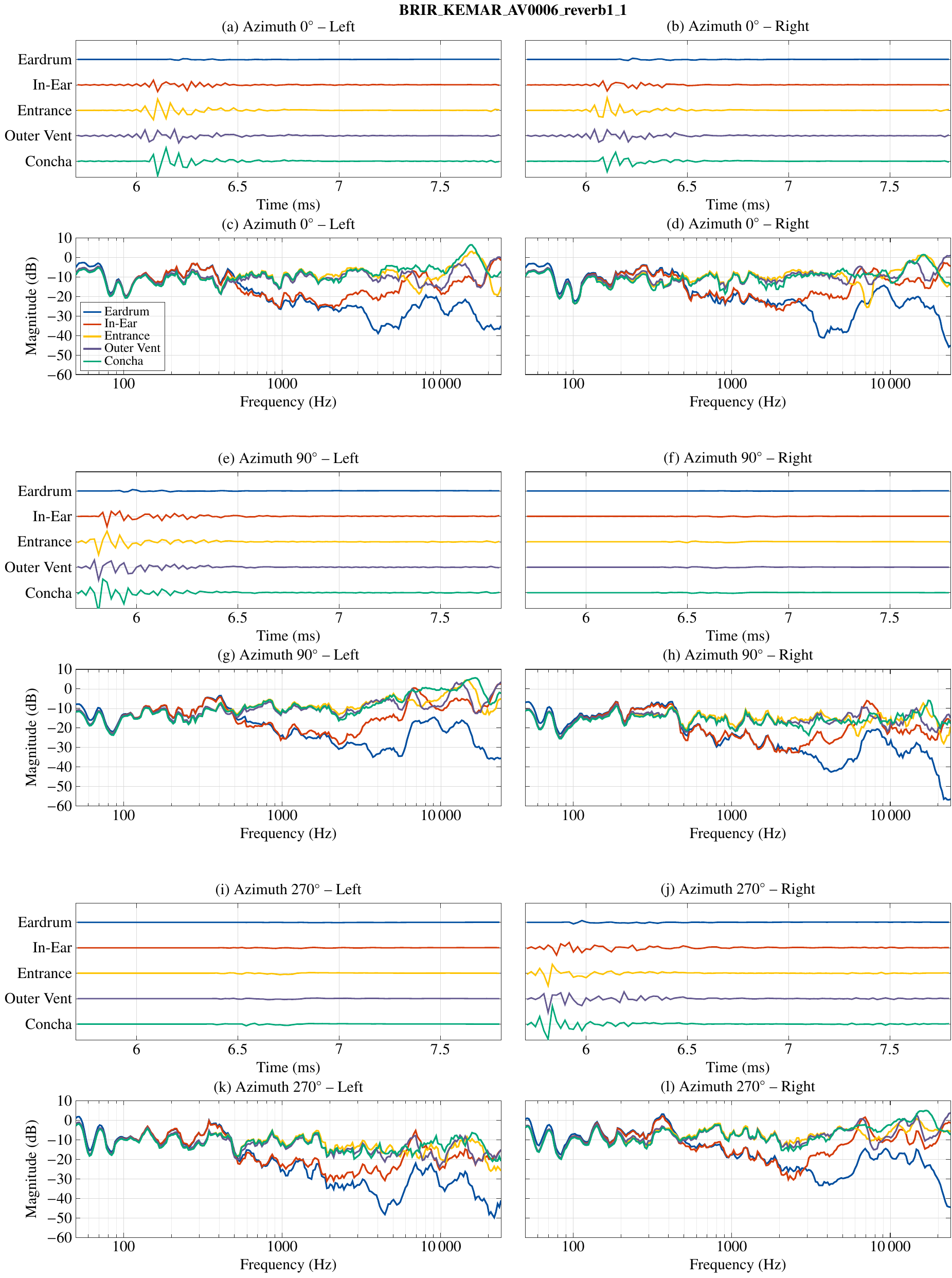}
    \vspace{-4pt}
    \caption{Representative BRIR data from \texttt{BRIR\_KEMAR\_AV0006\_reverb1\_1.mat} for source azimuths of \qty{0}{\degree}, \qty{90}{\degree}, and \qty{270}{\degree}. For each source direction, the time-domain responses of the 10 recording channels are shown above the corresponding magnitude responses, grouped by left and right side. The time-domain impulse responses are vertically offset for visualization.}
    \label{fig:brir-example}
\end{figure}

\Cref{fig:brir-example} shows representative BRIR data from Hearpiece configuration \texttt{AV0006} in room condition \texttt{reverb1}, insertion 1, for source azimuths of \qty{0}{\degree}, \qty{90}{\degree}, and \qty{270}{\degree}. For each source direction, the time-domain responses of all 10 recording channels are shown above their corresponding magnitude responses, grouped by left and right side. 

\FloatBarrier


\subsection{Driver responses}
\label{sec:driver-responses}

The driver-response data are stored in \texttt{DriverResponses\_MAT/}. The directory contains six MAT files corresponding to two Hearpiece configurations and three independent insertions. The files follow the naming convention
\begin{lstlisting}
DriverResponse_KEMAR_<deviceID>_<insertion>.mat
\end{lstlisting}
where \texttt{deviceID} is either \texttt{AC0031} or \texttt{AV0006}, and \texttt{insertion} is 1, 2, or 3. 
\vspace{1em}

The variables contained in each driver-response MAT file are summarized in \Cref{tab:driver-vars}. Each file contains the impulse responses in \texttt{M\_data} with dimensions $\num{1536} \times 4 \times 10$, corresponding to samples $\times$ drivers $\times$ recording channels. The entries of \texttt{c\_outchannel\_names} correspond to the driver dimension of \texttt{M\_data}, and the entries of \texttt{c\_inchannel\_names} correspond to the recording-channel dimension. The variable \texttt{n\_delay\_common} follows the timing convention used in the previous Hearpiece database \citep{denk2021hearpiece} and specifies the number of samples preceding the nominal time origin of the stored impulse responses. In the present database, \texttt{n\_delay\_common} is set to 0 for all driver-response files, indicating that no samples are designated as negative-time samples. Calibration and response corrections applied to the driver-response data are described in \Cref{sec:calibration}.

\begin{table}[ht]
\caption{Variables contained in each driver-response MAT file.}
\label{tab:driver-vars}
\centering
\begin{tabularx}{\linewidth}{@{}l l Y@{}}
\toprule
Variable & Size/type & Description \\
\midrule
\texttt{M\_data} & $1536 \times 4 \times 10$ & Impulse responses with dimensions of samples $\times$ drivers $\times$ recording channels. \\
\texttt{c\_inchannel\_names} & $1 \times 10$ cell array & Recording-channel names. \\
\texttt{c\_outchannel\_names} & $1 \times 4$ cell array & Driver names corresponding to the second dimension of \texttt{M\_data}. \\
\texttt{n\_delay\_common} & scalar & Number of negative-time samples preceding the nominal time origin; equal to 0. \\
\texttt{srate} & scalar & Sampling rate in Hz; equal to \num{48000}. \\
\texttt{s\_subjectID} & character array & Subject identifier; \texttt{KEMAR}. \\
\texttt{s\_deviceID} & character array & Hearpiece device identifier; \texttt{AC0031} or \texttt{AV0006}. \\
\texttt{s\_dataset} & character array & Dataset identifier. \\
\texttt{s\_reverbCondition} & character array & Room-condition identifier; \texttt{reverb1}. \\
\texttt{s\_fitting} & character array & Hearpiece fitting; \texttt{closed} or \texttt{vented}. \\
\texttt{n\_insertion} & scalar & Hearpiece insertion number. \\
\texttt{c\_readme} & cell array & Brief description of the data, measurement configuration, and processing. \\
\bottomrule
\end{tabularx}
\end{table}
\FloatBarrier

\begin{figure}[ht]
    \centering
    \includegraphics[width=0.96\linewidth]{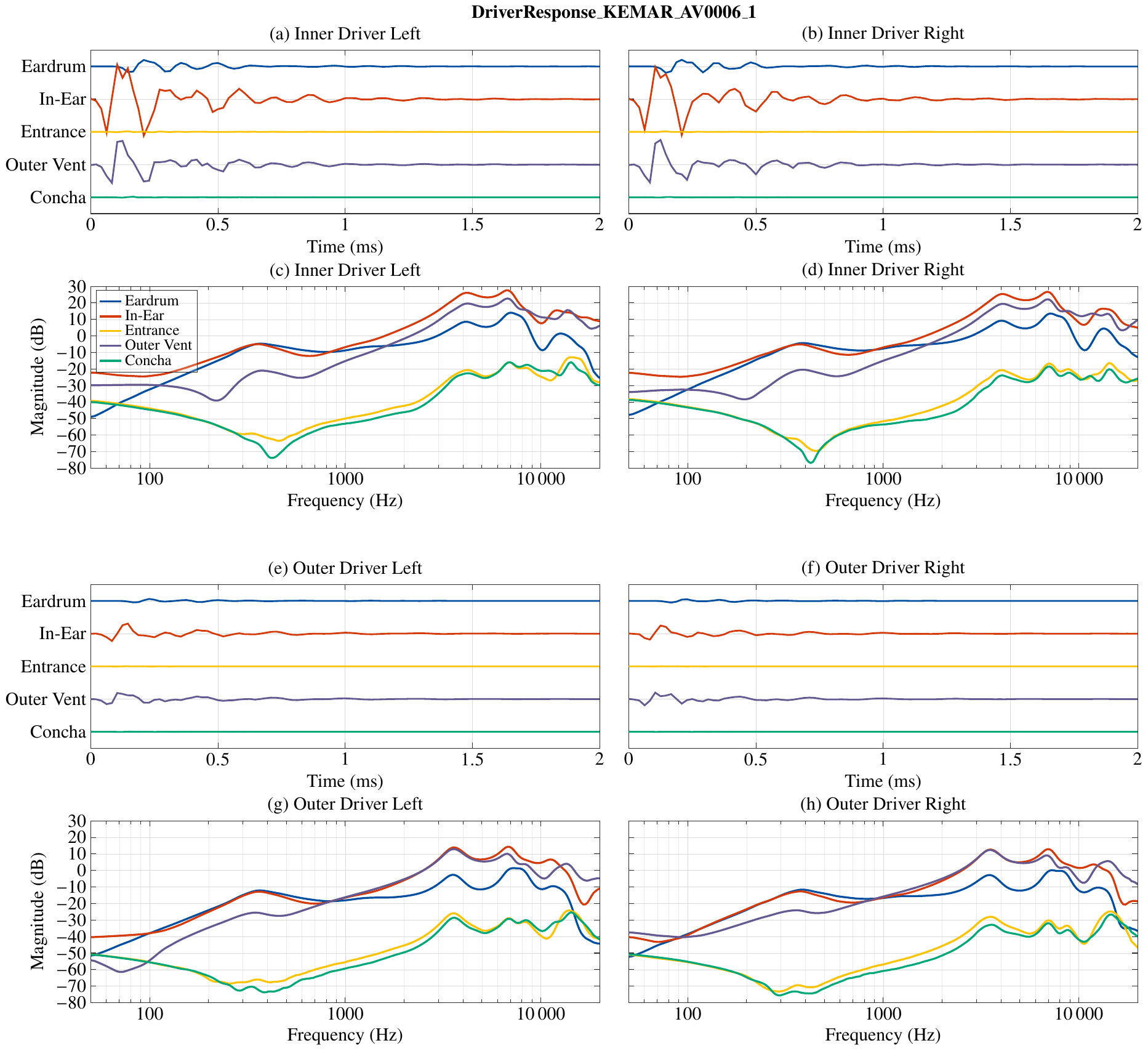}
    \vspace{-8pt}
    \caption{Representative driver-response data from \texttt{DriverResponse\_KEMAR\_AV0006\_1.mat}. Time-domain and corresponding magnitude responses are shown for Inner Driver Left, Inner Driver Right, Outer Driver Left, and Outer Driver Right. Each panel contains the five recording locations on the corresponding side. The time-domain impulse responses are vertically offset for visualization. }
    \label{fig:driver-example}
\end{figure}

\Cref{fig:driver-example} shows representative driver-response data from Hearpiece configuration \texttt{AV0006}, insertion 1. The time-domain panels show the responses from each of the four drivers to the five recording locations on the corresponding side: Eardrum, In-Ear, Entrance, Outer Vent, and Concha. The corresponding magnitude-response panels are shown directly below.
\FloatBarrier


\subsection{Center-microphone room impulse responses}
\label{sec:centermic}

Center-microphone RIRs are stored in \texttt{CenterMic\_MAT/}. One MAT file is provided for each of the three room conditions:
\begin{lstlisting}
CenterMic_reverb1.mat
CenterMic_reverb2.mat
CenterMic_reverb3.mat
\end{lstlisting}
The measurements were performed with KEMAR and the Hearpieces removed from the measurement position. A Sennheiser {KE 4-211-2} omnidirectional microphone was positioned at the center of the loudspeaker array, providing a reference characterization of the three room conditions for all 24 source directions.

The variables contained in each center-microphone MAT file are summarized in \Cref{tab:centermic-vars}. Each file contains the RIRs in \texttt{M\_data} with dimensions $\num{72000} \times 24$, corresponding to samples $\times$ source directions. The rows of \texttt{M\_dirs\_sph} correspond to the source-direction dimension of \texttt{M\_data}, and the entry of \texttt{c\_inchannel\_names} identifies the recording channel. The variable \texttt{n\_delay\_common} is set to 0 for all center-microphone files, indicating that no samples are designated as negative-time samples. Unlike the BRIR and driver-response data, no loopback compensation or microphone-response correction was applied to the center-microphone responses.

\begin{table}[ht]
\caption{Variables contained in each center-microphone MAT file.}
\label{tab:centermic-vars}
\centering
\begin{tabularx}{\linewidth}{@{}l l Y@{}}
\toprule
Variable & Size/type & Description \\
\midrule
\texttt{M\_data} & $\num{72000} \times 24$ & RIRs with dimensions of samples $\times$ source directions. \\
\texttt{M\_dirs\_sph} & $24 \times 2$ & Source directions given as azimuth and elevation in degrees. \\
\texttt{c\_inchannel\_names} & $1 \times 1$ cell array & Recording-channel name. \\
\texttt{n\_delay\_common} & scalar & Number of negative-time samples preceding the nominal time origin; equal to 0. \\
\texttt{srate} & scalar & Sampling rate in Hz; equal to \num{48000}. \\
\texttt{s\_dataset} & character array & Dataset identifier. \\
\texttt{s\_reverbCondition} & character array & Room-condition identifier; \texttt{reverb1}, \texttt{reverb2}, or \texttt{reverb3}. \\
\texttt{c\_readme} & cell array & Brief description of the data, measurement configuration, and processing. \\
\bottomrule
\end{tabularx}
\end{table}
\FloatBarrier

\begin{figure}[ht]
    \centering
    \includegraphics[width=\linewidth]{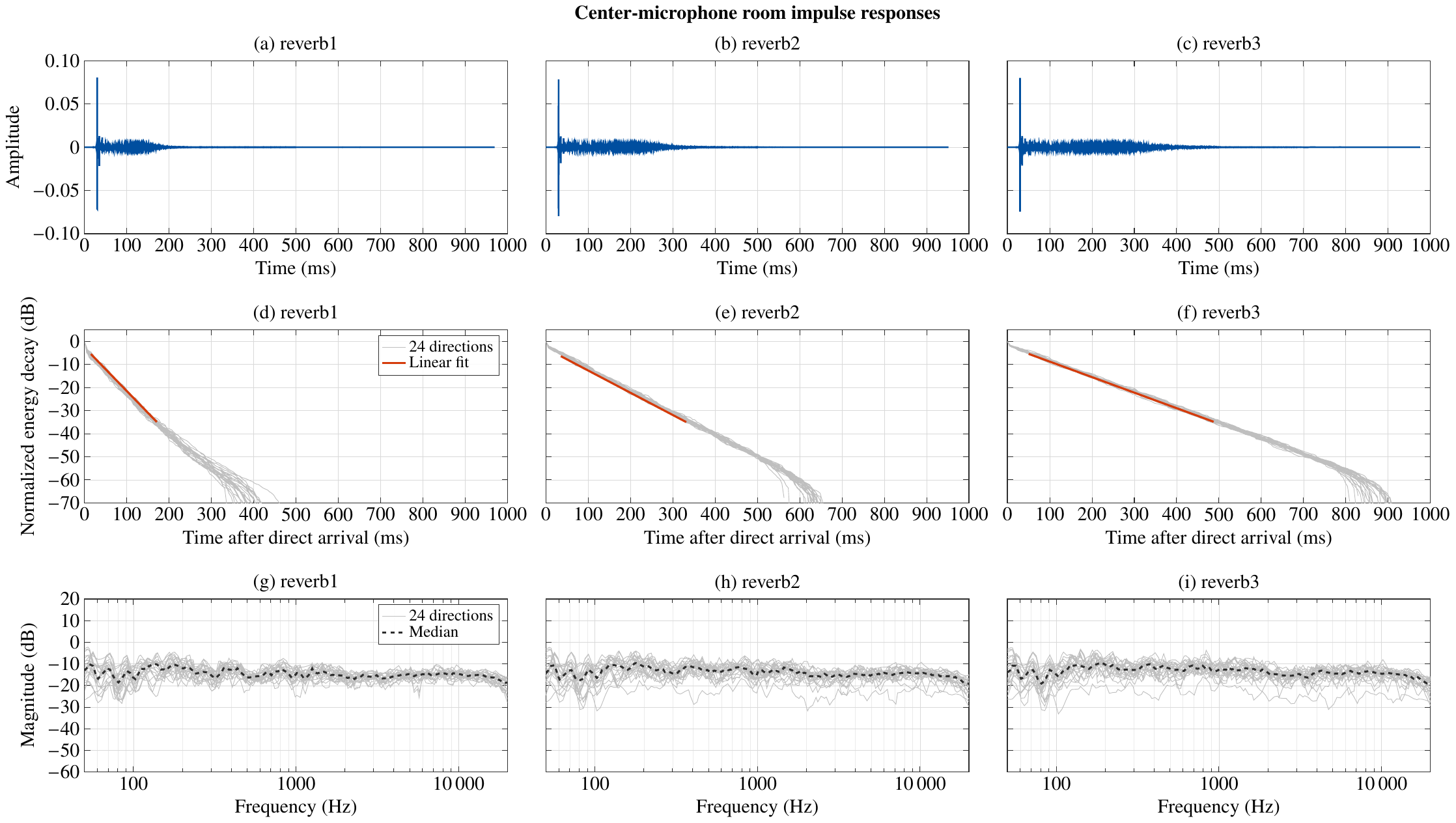}
    \vspace{-14pt}
    \caption{Center-microphone responses for the three room conditions. Top: the RIR for the source direction corresponding to loudspeaker \texttt{\#2} in each room condition. Middle: normalized \qty{1}{\kilo\hertz} energy decay curves for all 24 source directions and a linear decay fit between \qty{-5}{\decibel} and \qty{-35}{\decibel}. Bottom: magnitude responses for all 24 source directions and their median.}
    \label{fig:centermic-example}
    \vspace{12pt}
\end{figure}

\Cref{fig:centermic-example} illustrates the center-microphone responses for the three room conditions. The top row shows an example RIR for the same source direction in each room condition. The middle row shows normalized energy decay curves in the \qty{1}{\kilo\hertz} octave band for all 24 source directions. The decay curves are referenced to the detected direct arrival, and the linear decay fit is evaluated between \qty{-5}{\decibel} and \qty{-35}{\decibel}. The reverberation time is estimated by extrapolating the fitted decay slope to a \qty{60}{\decibel} decay. The median estimated reverberation times across the 24 source directions are approximately \qty{0.32}{\second}, \qty{0.62}{\second}, and \qty{0.89}{\second} for \texttt{reverb1}, \texttt{reverb2}, and \texttt{reverb3}, respectively. The bottom row shows the magnitude responses for all 24 source directions together with their median.

\FloatBarrier


\subsection{Loopback measurement}
\label{sec:loopback-measurement}

The measured loopback impulse response is stored in \texttt{Calibration\_MAT/Loopback.mat}. The variables contained in the loopback MAT file are summarized in \Cref{tab:loopback-vars}. The file contains the measured loopback impulse response in \texttt{M\_data} with dimensions $\num{334227} \times 1$, corresponding to samples $\times$ recording channels. The loopback impulse response was used to compensate the BRIR and driver-response data for the response of the playback and recording system. No loopback compensation was applied to the center-microphone responses.

\begin{table}[ht]
\caption{Variables contained in the loopback MAT file.}
\label{tab:loopback-vars}
\centering
\begin{tabularx}{\linewidth}{@{}l l Y@{}}
\toprule
Variable & Size/type & Description \\
\midrule
\texttt{M\_data} & $\num{334227} \times 1$ & Measured single-channel loopback impulse response. \\
\texttt{srate} & scalar & Sampling rate in Hz; equal to \num{48000}. \\
\texttt{s\_dataset} & character array & Dataset identifier. \\
\texttt{s\_calibration} & character array & Calibration-measurement identifier. \\
\texttt{s\_dataType} & character array & Description of the stored calibration data. \\
\texttt{c\_readme} & cell array & Brief description of the data, measurement configuration, and processing. \\
\bottomrule
\end{tabularx}
\end{table}
\FloatBarrier

\begin{figure}[ht]
    \centering
    \includegraphics[width=0.56\linewidth]{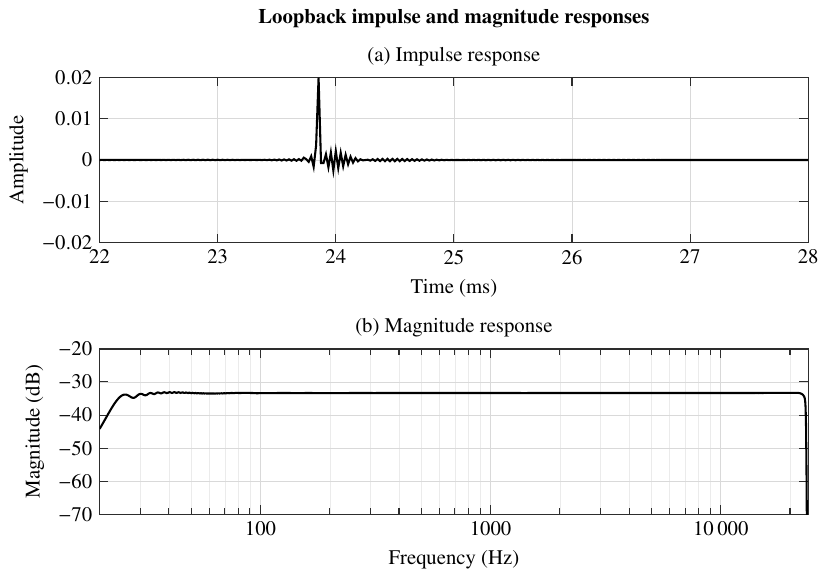}
    \vspace{-8pt}
    \caption{Measured loopback impulse response and corresponding magnitude response of the playback and recording system. Top: loopback impulse response from \qty{22}{\milli\second} to \qty{28}{\milli\second}. Bottom: magnitude response computed from the full loopback impulse response over the frequency range from \qty{20}{\hertz} to \qty{24}{\kilo\hertz}.}
    \label{fig:loopback-example}
    \vspace{-10pt}
\end{figure}

\Cref{fig:loopback-example} shows the loopback impulse response and the corresponding magnitude response. The time-domain panel shows the interval from \qty{22}{\milli\second} to \qty{28}{\milli\second}, containing the main impulse response. The frequency-domain panel shows the magnitude response computed from the full loopback impulse response over the frequency range from \qty{20}{\hertz} to \qty{24}{\kilo\hertz}.

\FloatBarrier

\section{Example applications}
\label{sec:applications}

The BRIR data can be used to simulate the signals at the Eardrum locations and Hearpiece microphones for sound arriving from different source directions and under different room conditions. This enables evaluation of microphone-array signal-processing algorithms such as beamforming, source localization, and source separation, as well as related speech-enhancement algorithms \citep{doclo2015multichannel,gannot2017consolidated,wang2018supervised}. For active sound-control applications, the BRIRs can additionally represent the propagation paths from external sources to the Hearpiece microphones and Eardrum locations.

The driver-response data characterize the acoustic transfer paths from each Hearpiece driver to the Eardrum locations and Hearpiece microphones. In active noise control (ANC), the responses from the Hearpiece drivers to the In-Ear microphones and Eardrum locations can serve as secondary paths for controller design, virtual sensing, and robustness analysis \citep{benois2021eardrum,benois2022virtual,hilgemann2024uncertainty,shi2022sfanc,luo2025gfanc,yang2025meta,luo2026rlsfanc}. Spatially selective active noise control (SSANC) combines spatial filtering with active control to preserve a desired source while selectively attenuating interfering sources \citep{xiao2023spatial,xiao2024icassp,xiao2025fa,xiao2025soft,xiao2026robust}. The same driver-to-In-Ear and driver-to-Eardrum responses are also relevant to active occlusion control (AOC) \citep{liebich2018aoc,liebich2022occlusion}. In acoustic feedback cancellation (AFC), the responses from the Hearpiece drivers to the Entrance, Outer Vent, and Concha microphones can be regarded as acoustic feedback paths when these microphones are used as device input microphones \citep{schepker2019null,schepker2020soft,soleimani2023neural,lydaki2026deep,voit2026intheloop,zhan2025deeppem,hao2025l3c}.

The BRIR and driver-response data can also be used jointly to construct linear acoustic simulations that include both external-source propagation and sound generated by the Hearpiece drivers, for example in ANC, SSANC, and AFC applications. The two Hearpiece configurations and three independent insertions further allow such algorithms to be evaluated with configuration and insertion variability.

\bibliographystyle{IEEEtranN}
\bibliography{References}

@inproceedings{denk2019earpiece,
  author    = {Denk, Florian and Lettau, Miriam and Schepker, Henning and Doclo, Simon and Roden, Reinhild and Blau, Matthias and Bach, J{\"o}rg-Hendrik and Wellmann, Jan and Kollmeier, Birger},
  title     = {A one-size-fits-all earpiece with multiple microphones and drivers for hearing device research},
  booktitle = {Proc. AES International Conference on Headphone Technology},
  year      = {2019},
  month     = aug,
  address   = {San Francisco, CA, USA},
  pages     = {1--9},
  note      = {{Paper} 13},
}

@manual{hearpiece2019technical,
  author = {{H{\"o}rTech gGmbH} and {{Carl von Ossietzky Universit{\"a}t Oldenburg}} and {{InEar GmbH}}},
  title  = {Technical Documentation: Acoustically Transparent Earpieces},
  year   = {2019},
  url    = {https://api.inear.de/user/pages/04.support/03.downloads/02._group/04._downloads/InEar%20%26%20H%C3%B6rTech%20Transparent%20EarPiece.pdf}
}

@article{denk2021hearpiece,
  author  = {Denk, Florian and Kollmeier, Birger},
  title   = {The {Hearpiece} database of individual transfer functions of an in-the-ear earpiece for hearing device research},
  journal = {Acta Acustica},
  year    = {2021},
  volume  = {5},
  note    = {\relax{Art. no. 2}},
  pages   = {1--16},
  doi     = {10.1051/aacus/2020028}
}

@misc{ohlenbusch2024transfer,
  author    = {Ohlenbusch, Mattes and Rollwage, Christian and Doclo, Simon},
  title     = {Transfer Function Measurements for Simulating Environmental Noise at Hearable Microphones},
  year      = {2024},
  publisher = {Zenodo},
  note      = {Version 1.0},
  doi       = {10.5281/zenodo.11196867},
  url       = {https://doi.org/10.5281/zenodo.11196867}
}

@inproceedings{liebich2019database,
  author    = {Liebich, Stefan and Fabry, Johannes and Jax, Peter and Vary, Peter},
  title     = {Acoustic Path Database for {ANC} In-Ear Headphone Development},
  booktitle = {Proc. 23rd International Congress on Acoustics (ICA)},
  pages     = {4326--4333},
  address   = {Aachen, Germany},
  month     = sep,
  year      = {2019},
  doi       = {10.18154/RWTH-CONV-239404}
}

@article{kayser2009database,
  author  = {Kayser, Hendrik and Ewert, Stephan D. and Anem{\"u}ller, J{\"o}rn and Rohdenburg, Tim and Hohmann, Volker and Kollmeier, Birger},
  title   = {Database of Multichannel In-Ear and Behind-the-Ear Head-Related and Binaural Room Impulse Responses},
  journal = {EURASIP Journal on Advances in Signal Processing},
  volume  = {2009},
  pages   = {1--10},
  note    = {\relax{Art. no. 298605}},
  year    = {2009},
  doi     = {10.1155/2009/298605}
}

@inproceedings{fejgin2023brudex,
  author    = {Fejgin, Daniel and Middelberg, Wiebke and Doclo, Simon},
  title     = {{BRUDEX} Database: Binaural Room Impulse Responses with Uniformly Distributed External Microphones},
  booktitle = {Proc. ITG Conference on Speech Communication},
  pages     = {1--5},
  address   = {Aachen, Germany},
  month     = sep,
  year      = {2023},
  doi       = {10.30420/456164024}
}

@misc{pausch2022ihta,
  author    = {Pausch, Florian and Doma, Shaima'a and Fels, Janina},
  title     = {{IHTA-indHARTF} -- Database of Individual Behind-the-Ear Hearing-Aid-Related Transfer Functions with High Spatial Resolution},
  year      = {2022},
  publisher = {RWTH Aachen University},
  doi       = {10.18154/RWTH-2022-04267},
  url       = {https://doi.org/10.18154/RWTH-2022-04267}
}

@manual{knowlesSPH1642HT5H1,
  author       = {{Knowles Electronics}},
  title        = {{SPH1642HT5H-1: Wide Bandwidth, Low Noise, Precision Top Port SiSonic Microphone}},
  year         = {2015},
  organization = {Knowles Electronics},
  note         = {Rev. B},
  url          = {https://www.digikey.com/en/products/detail/syntiant/SPH1642HT5H-1/5805540}
}

@misc{grasCalibration,
  author = {{G.R.A.S. Sound \& Vibration}},
  title  = {{Certificate of Calibration: 45BB-12 KEMAR Head \& Torso with Anthropometric Pinnae and 2-Channel Low Noise Ear Simulator}},
  year   = {2018},
  note   = {{KEMAR} serial no. 318190, calibration date: 21 August 2018}
}

@inproceedings{farina2000swept,
  author    = {Farina, Angelo},
  title     = {Simultaneous Measurement of Impulse Response and Distortion with a Swept-Sine Technique},
  booktitle = {Proc. 108th Audio Engineering Society Convention},
  year      = {2000},
  month     = feb,
  address   = {Paris, France},
  note      = {{Paper} 5093}
}

@article{novak2010nonlinear,
  author  = {Nov{\'a}k, Antonin and Simon, Laurent and Kadlec, Franti{\v{s}}ek and Lotton, Pierrick},
  title   = {Nonlinear System Identification Using Exponential Swept-Sine Signal},
  journal = {IEEE Transactions on Instrumentation and Measurement},
  year    = {2010},
  volume  = {59},
  number  = {8},
  pages   = {2220--2229},
  doi     = {10.1109/TIM.2009.2031836}
}

@article{doclo2015multichannel,
  author  = {Doclo, Simon and Kellermann, Walter and Makino, Shoji and Nordholm, Sven Erik},
  title   = {Multichannel signal enhancement algorithms for assisted listening devices: Exploiting spatial diversity using multiple microphones},
  journal = {IEEE Signal Processing Magazine},
  volume  = {32},
  number  = {2},
  pages   = {18--30},
  year    = {2015}
}

@article{gannot2017consolidated,
  author  = {Gannot, Sharon and Vincent, Emmanuel and Markovich-Golan, Shmulik and Ozerov, Alexey},
  title   = {A consolidated perspective on multi-microphone speech enhancement and source separation},
  journal = {IEEE/ACM Transactions on Audio, Speech, and Language Processing},
  volume  = {25},
  number  = {4},
  pages   = {692--730},
  year    = {2017}
}

@article{wang2018supervised,
  author  = {DeLiang Wang and Jitong Chen},
  title   = {Supervised speech separation based on deep learning: An overview},
  journal = {IEEE/ACM Transactions on Audio, Speech, and Language Processing},
  volume  = {26},
  number  = {10},
  pages   = {1702--1726},
  year    = {2018},
  doi     = {10.1109/TASLP.2018.2842159}
}

@inproceedings{benois2021eardrum,
  author    = {Rivera Benois, Piero and Roden, Reinhild and Blau, Matthias and Doclo, Simon},
  title     = {Sound Pressure Minimization at the Ear Drum for In-Ear {ANC} Headphones Using a Fixed Feedforward Remote Microphone Technique},
  booktitle = {Proc. 27th International Congress on Sound and Vibration (ICSV27)},
  pages     = {473--480},
  address   = {{Online}},
  month     = jul,
  year      = {2021}
}

@inproceedings{benois2022virtual,
  author    = {Rivera Benois, Piero and Roden, Reinhild and Blau, Matthias and Doclo, Simon},
  title     = {Optimization of a Fixed Virtual Sensing Feedback {ANC} Controller for In-Ear Headphones with Multiple Loudspeakers},
  booktitle = {Proc. IEEE International Conference on Acoustics, Speech and Signal Processing (ICASSP)},
  pages     = {8717--8721},
  year      = {2022},
  doi       = {10.1109/ICASSP43922.2022.9746327}
}

@article{hilgemann2024uncertainty,
  author  = {Hilgemann, Florian and Chatzimoustafa, Egke and Jax, Peter},
  title   = {Data-Driven Uncertainty Modeling for Robust Feedback Active Noise Control in Headphones},
  journal = {Journal of the Audio Engineering Society},
  volume  = {72},
  number  = {12},
  pages   = {873--883},
  year    = {2024},
  doi     = {10.17743/jaes.2022.0185}
}

@article{shi2022sfanc,
  author  = {Shi, Dongyuan and Lam, Bhan and Ooi, Kenneth and Shen, Xiaoyi and Gan, Woon-Seng},
  title   = {Selective Fixed-Filter Active Noise Control Based on Convolutional Neural Network},
  journal = {Signal Processing},
  volume  = {190},
  pages   = {1--7},
  year    = {2022},
  note    = {\relax{Art. no. 108317}},
  doi     = {10.1016/j.sigpro.2021.108317}
}

@article{luo2025gfanc,
  author  = {Luo, Zhengding and Ji, Junwei and Wang, Boxiang and Shi, Dongyuan and Ma, Haozhe and Gan, Woon-Seng},
  title   = {Deep Learning-Based Generative Fixed-Filter Active Noise Control: Transferability and Implementation},
  journal = {Mechanical Systems and Signal Processing},
  volume  = {238},
  pages   = {1--17},
  year    = {2025},
  note    = {\relax{Art. no. 113207}},
  doi     = {10.1016/j.ymssp.2025.113207}
}

@inproceedings{yang2025meta,
  author    = {Yang, Ziyi and Luo, Zhengding and Shi, Dongyuan and Ji, Junwei and Wang, Boxiang and Li, Haowen and Huang, Qirui and Gan, Woon-Seng},
  title     = {Meta-Learned Regional Initialization of Control Filters for Headphone Active Noise Control},
  booktitle = {Proc. Asia Pacific Signal and Information Processing Association Annual Summit and Conference (APSIPA ASC)},
  pages     = {441--446},
  year      = {2025},
  doi       = {10.1109/APSIPAASC65261.2025.11249354}
}

@article{luo2026rlsfanc,
  author  = {Luo, Zhengding and Ma, Haozhe and Wang, Boxiang and Shi, Dongyuan and Gan, Woon-Seng},
  title   = {Reinforcement Learning-Based Selective Fixed-Filter Active Noise Control ({RL-SFANC}): From Theory to Real-Time Headphone Implementation},
  journal = {Signal Processing},
  volume  = {248},
  pages   = {1--12},
  year    = {2026},
  note    = {\relax{Art. no. 110695}},
  doi     = {10.1016/j.sigpro.2026.110695}
}

@article{xiao2023spatial,
  author  = {Xiao, Tong and Xu, Buye and Zhao, Chuming},
  title   = {Spatially Selective Active Noise Control Systems},
  journal = {The Journal of the Acoustical Society of America},
  volume  = {153},
  number  = {5},
  pages   = {2733--2744},
  year    = {2023},
  doi     = {10.1121/10.0019336}
}

@inproceedings{xiao2024icassp,
  author    = {Xiao, Tong and Doclo, Simon},
  title     = {Effect of Target Signals and Delays on Spatially Selective Active Noise Control for Open-Fitting Hearables},
  booktitle = {Proc. IEEE International Conference on Acoustics, Speech and Signal Processing (ICASSP)},
  pages     = {1056--1060},
  year      = {2024},
  doi       = {10.1109/ICASSP48485.2024.10445843}
}

@inproceedings{xiao2025fa,
  author    = {Xiao, Tong and Doclo, Simon},
  title     = {Spatially Selective Active Noise Control for Open-Fitting Hearables with Acausal Optimization},
  booktitle = {Proc. Forum Acusticum/Euronoise},
  address   = {M\'{a}laga, Spain},
  pages     = {117--124},
  month     = jun,
  year      = {2025},
  doi       = {10.61782/fa.2025.0817}
}

@inproceedings{xiao2025soft,
  author    = {Xiao, Tong and Roden, Reinhild and Blau, Matthias and Doclo, Simon},
  title     = {Soft-Constrained Spatially Selective Active Noise Control for Open-Fitting Hearables},
  booktitle = {Proc. IEEE Workshop on Applications of Signal Processing to Audio and Acoustics (WASPAA)},
  pages     = {1--5},
  year      = {2025},
  doi       = {10.1109/WASPAA66052.2025.11231002}
}

@inproceedings{xiao2026robust,
  author    = {Xiao, Tong and Roden, Reinhild and Blau, Matthias and Doclo, Simon},
  title     = {Robust Soft-Constrained Spatially Selective Active Noise Control for Hearables Under Secondary Path Variations},
  booktitle = {Proc. International Workshop on Acoustic Signal Enhancement (IWAENC)},
  address   = {Cremona, Italy},
  pages     = {1--5},
  month     = sep,
  year      = {2026}
}

@inproceedings{liebich2018aoc,
  author    = {Liebich, Stefan and Brandis, Raphael and Fabry, Johannes and Jax, Peter and Vary, Peter},
  title     = {Active Occlusion Cancellation with Hear-Through Equalization for Headphones},
  booktitle = {Proc. IEEE International Conference on Acoustics, Speech and Signal Processing (ICASSP)},
  pages     = {241--245},
  year      = {2018},
  doi       = {10.1109/ICASSP.2018.8461834}
}

@article{liebich2022occlusion,
  author  = {Liebich, Stefan and Vary, Peter},
  title   = {Occlusion Effect Cancellation in Headphones and Hearing Devices---The Sister of Active Noise Cancellation},
  journal = {IEEE/ACM Transactions on Audio, Speech, and Language Processing},
  volume  = {30},
  pages   = {35--48},
  year    = {2022},
  doi     = {10.1109/TASLP.2021.3130966}
}

@article{schepker2019null,
  author  = {Schepker, Henning and Nordholm, Sven Erik and Tran, Linh Thi Thuc and Doclo, Simon},
  title   = {Null-Steering Beamformer-Based Feedback Cancellation for Multi-Microphone Hearing Aids with Incoming Signal Preservation},
  journal = {IEEE/ACM Transactions on Audio, Speech, and Language Processing},
  volume  = {27},
  number  = {4},
  pages   = {679--691},
  year    = {2019},
  doi     = {10.1109/TASLP.2019.2892234}
}

@article{schepker2020soft,
  author  = {Schepker, Henning and Nordholm, Sven and Doclo, Simon},
  title   = {Acoustic Feedback Suppression for Multi-Microphone Hearing Devices Using a Soft-Constrained Null-Steering Beamformer},
  journal = {IEEE/ACM Transactions on Audio, Speech, and Language Processing},
  volume  = {28},
  pages   = {929--940},
  year    = {2020},
  doi     = {10.1109/TASLP.2020.2975390}
}

@inproceedings{soleimani2023neural,
  author    = {Soleimani, Behrad and Schepker, Henning and Mirbagheri, Majid},
  title     = {{Neural-AFC}: Learning-Based Step-Size Control for Adaptive Feedback Cancellation with Closed-Loop Model Training},
  booktitle = {Proc. IEEE International Conference on Acoustics, Speech and Signal Processing (ICASSP)},
  pages     = {1--5},
  year      = {2023},
  doi       = {10.1109/ICASSP49357.2023.10094841}
}

@article{lydaki2026deep,
  author  = {Lydaki, Eleftheria and Tan, Zheng-Hua and Jensen, Jesper and Guo, Meng},
  title   = {Deep Feedback Cancellation in Hearing Aids},
  journal = {IEEE Transactions on Audio, Speech and Language Processing},
  volume  = {34},
  pages   = {3159--3173},
  year    = {2026},
  doi     = {10.1109/TASLPRO.2026.3700049}
}

@inproceedings{voit2026intheloop,
  author    = {Voit, Svantje and Doclo, Simon},
  title     = {In-the-Loop Training of Deep Feedback Cancellation for Hearing Aids},
  booktitle = {Proc. International Workshop on Acoustic Signal Enhancement (IWAENC)},
  address   = {Cremona, Italy},
  pages     = {1--5},
  month     = sep,
  year      = {2026}
}

@inproceedings{zhan2025deeppem,
  author    = {Zhan, Xiaofan and Hao, Fengyuan and Li, Xiaodong and Zheng, Chengshi},
  title     = {{DeepPEM-AFC}: An Improved Prediction-Error-Method-Based Adaptive Feedback Cancellation with Deep Learning for Hearing Aids},
  booktitle = {Proc. IEEE International Conference on Acoustics, Speech and Signal Processing (ICASSP)},
  pages     = {1--5},
  year      = {2025},
  doi       = {10.1109/ICASSP49660.2025.10890348}
}

@inproceedings{hao2025l3c,
  author    = {Hao, Fengyuan and Moore, Brian C. J. and Zhang, Huiyong and Li, Xiaodong and Zheng, Chengshi},
  title     = {{L3C-DeepMFC}: Low-Latency Low-Complexity Deep Marginal Feedback Cancellation with Closed-Loop Fine Tuning for Hearing Aids},
  booktitle = {Proc. Interspeech},
  pages     = {838--842},
  year      = {2025},
  doi       = {10.21437/Interspeech.2025-1111}
}

\end{document}